\documentclass[11pt,twoside]{article}  
\usepackage{adassconf}
\usepackage{psfig}

\begin{document}   

%
%

\paperID{P2.2.7}

%
%
%
%

\title{Time Domain Explorations With Digital Sky Surveys}

%

\author{Ashish A.\ Mahabal, S.\ G.\ Djorgovski, M.\ J.\ Graham, Priya Kollipara, 
Benjamin Granett, Elisabeth Krause, Roy Williams, M.\ Bogosavljevic}
\affil{California Institute of Technology, Pasadena, CA 91125}
\author{C.\ Baltay, D.\ Rabinowitz, A.\ Bauer, P.\ Andrews, N.\ Ellman,
S.\ Duffau, J.\ Jerke}\affil{Yale University, New Haven, CT 06520}
\author{A.\ Rengstorf, R.\ Brunner} \affil{NCSA/UIUC, Champaign, IL 61820}
\author{J.\ Musser, S.\ Mufson, M.\ Gebhard} \affil{Indiana University,
Bloomington, IN 47405}



\contact{Ashish A.\ Mahabal}
\email{aam@astro.caltech.edu}

%
%
%
%
%

\paindex{Mahabal, A. A.}
\aindex{Djorgovski, S. G.}
\aindex{Williams, R.}
\aindex{Graham, M. J.}
\aindex{Kollipara, P.}
\aindex{Bogosavljevic, M.}
\aindex{Krause, E.}
\aindex{Granett, B.}
\aindex{Baltay, C.}
\aindex{Rabinowitz, D.}
\aindex{Bauer, A.}
\aindex{Andrews, P.}
\aindex{Ellman, N.}
\aindex{Duffau, S.}
\aindex{Snyder, J.}
\aindex{Morgan, N.}
\aindex{Rengstorf, A.}
\aindex{Brunner, R.}
\aindex{Musser, J.}
\aindex{Mufson, S.}
\aindex{Gebhard, M.} 

%
%

\authormark{Mahabal et al.}




\begin{abstract}          
One of the new frontiers of astronomical research is the exploration
of time variability on the sky at different wavelengths and flux levels.
We have carried out a pilot project using DPOSS data to study strong
variables and transients, and are now extending it to the new
Palomar-QUEST synoptic sky survey. We report on our early findings
and outline the methodology to be implemented in preparation for a
real-time transient detection pipeline. In addition to large numbers
of known types of highly variable sources (e.g., SNe, CVs, OVV QSOs,
etc.), we expect to find numerous transients whose nature may be
established by a rapid follow-up.  Whereas we will make all detected
variables publicly available through the web, we anticipate that email
alerts would be issued in the real time for a subset of events deemed
to be the most interesting.  This real-time process entails many
challenges, in an effort to maintain a high completeness while keeping
the contamination low.  We will utilize distributed Grid services
developed by the GRIST project, and implement a variety of advanced
statistical and machine learning techniques.
\end{abstract}


\section{Introduction}

Systematic exploration of
previously poorly covered regions of the observable parameter space
is a major source of discoveries in astronomy
(Djorgovski et al. 2001a, 2001b).
In particular, exploration of the time variability on the sky over a broad
range of flux levels and wavelengths is rapidly becoming a new frontier
of astronomical research (Paczynski 2000, Diercks 2001).
All manner of variable stars,
stellar explosions such as SNe and GRBs,
variable AGN, pulsars, microlensing events,  etc. are some of the examples
of currently known exciting time domain astrophysical phenomena.
Many more are certain to be discovered over time as planned ambitious
synoptic sky surveys (Tyson 2002, Kaiser 2002) join the existing ones
involving the time domain (see, e.g., Paczynski 2001,
http://www.astro.princeton.edu/$\sim$bp/ for listings).
The important factors
in such programs are (1) the area covered, (2) the depth of coverage, (3)
number of wavelengths used, and (4) the baseline(s) in time.

We briefly describe here a pilot project we carried out with
the plate overlap regions of the Digital Palomar Observatory Sky Survey
(hereafter DPOSS). We then outline ongoing programs
with the Palomar-QUEST sky survey (hereafter PQ) and describe
a realtime transient detection system and its challenges.

\section{Variability with DPOSS overlap regions}
DPOSS covers the entire
Northern Hemisphere in three filters viz. J, F and N plates (calibrated to
Gunn {\it g, r, i}). Each plate
is 6.5 degrees wide with adjacent plates overlapping by strips that are 1.5 degrees
wide. This results in at least 40\% of the sky being imaged at least twice
in each of the three filters.
We conducted an exploratory search for highly variable objects (Granett et al. 2005)
and optical transients (Mahabal et al. 2005) using
$\sim 8000~deg^2$ in the NGP and SGP areas of these plate overlap regions.
The effective depth of these searches was $r_{max} \approx 19$ mag for the
``high'' states, with a plate limits $r_{max} \approx 21$ mag.
Time baselines ranged from days to years, with $\sim 2$ yrs being
typical.
After eliminating various artifacts and contaminants, and applying well
defined statistical criteria for selection, we identified a large number of
highly variable objects, and followed up spectroscopically a subset of them
at the Palomar 200-inch telescope.  They turned out to be a heterogeneous
collection of flaring M-dwarfs, OVV QSOs and BL Lacs, CVs (including a rare
magnetic one), and some otherwise non-descript stars.  Approximately a
third to a half of all highly variable objects down to this magnitude level,
at moderate and high Galactic latitudes appear to be associated with AGN.

We also found a number of optical transients (operationally defined as
high-S/N, PSF-like objects, detected only once).
We estimate that a single-epoch ``snapshot'' down to this flux level contains
up to $\sim 1000$ transients/sky.
Their nature remains unknown, but in at least 2 cases deep follow-up imaging
revealed apparent faint host galaxies, which now await spectroscopy.

This pilot study gave us some hints as to what may be expected in a dedicated,
wide-field, synoptic sky survey at comparable magnitudes.  The faint variable
sky has a very rich and diverse phenomenology.

\section{The driftscans of Palomar-QUEST}
The Palomar-Quest synoptic sky survey (Djorgovski et al. 2004, 2005;
Baltay et al. 2005),
a collaborative project between Yale, Caltech,
and NCSA (some other groups are also involved in more specific collaborations)
is a new major digital sky survey conducted at the Samuel Oschin 48-inch
Schmidt telescope at Palomar.
The survey uses a special 112-CCD, 162-Megapixel camera built especially for
this purpose.  Some of the salient features of the survey are:
(1) Data taking in the Point-and Stare (PS) mode, covering $\sim 9.2 ~deg^2$
per exposure, or in a Drift Scan (DS) mode, in strips $4.6^{\rm o}$ wide, with a
typical coverage of $\sim 500 ~deg^2$/night,
(2) Near simultaneous observations in one of two filter sets in the DS mode:
Johnson-Cousins {\it UBRI} or SDSS {\it r'i'z'z'},
(3) In good conditions, typical limiting magnitudes for point sources:
$R_{lim} \approx 22$ mag, $I_{lim} \approx 21$ mag,
(4) In the DS mode, useful Declination range
$-25^{\rm o} < \delta < +30^{\rm o}$, for a
total anticipated survey area of $\sim 15,000 ~deg^2$,
(5) Multiple-pass coverage, with at least 4 passes per year at each covered
location,
(6) Time baselines for repeats ranging from days to months, anticipated to
extend to multi-year time scales over the next 3 to 5 years or beyond,
(7) NVO standards, protocols, and connections built in from the start.

The survey has started producing a steady stream of science-grade data,
from summer of 2003.  In the DS mode, it typically generates $\sim 1$ TB of
raw image data per month (assuming $\sim 14$ clear nights).
This unprecedented amount of data makes this the largest synoptic survey of its
kind both in terms of area covered and depth.  A broad range of science is
envisioned for the survey, but exploration of the time domain will be one of
the main focal areas.

PQ coverage as of Oct. 2004 is as follows: 
$\sim11500 ~deg ^2$ have been covered 
in {\it UBRI} ($\sim8900 ~deg^2$ at least twice) and $\sim 12100~deg^2$ in {\it r'i'z'z'}
($\sim9300 ~deg^2$ at least twice).  We have been testing our transient
detection techniques on areas that have been observed a large number of times.
These techniques are being perfected in preparation of a realtime pipeline
described in the next section.
Besides the more exciting transient detections is also science which involves
objects that simply vary in interesting ways. 
A large number of epochs with a range of baselines for several tens of
thousands of quasars will provide good limits to differentiate between the models
and further lead to better estimates of quasar lifetimes. This is possible by combining
PQ data with earlier surveys like SDSS, DPOSS, DSS etc.
Figure 1 shows B-band structure function for 500 quasars with $1 < z < 4$, a
starting point for such a study.

\begin{figure} 
\centerline{
\psfig{file=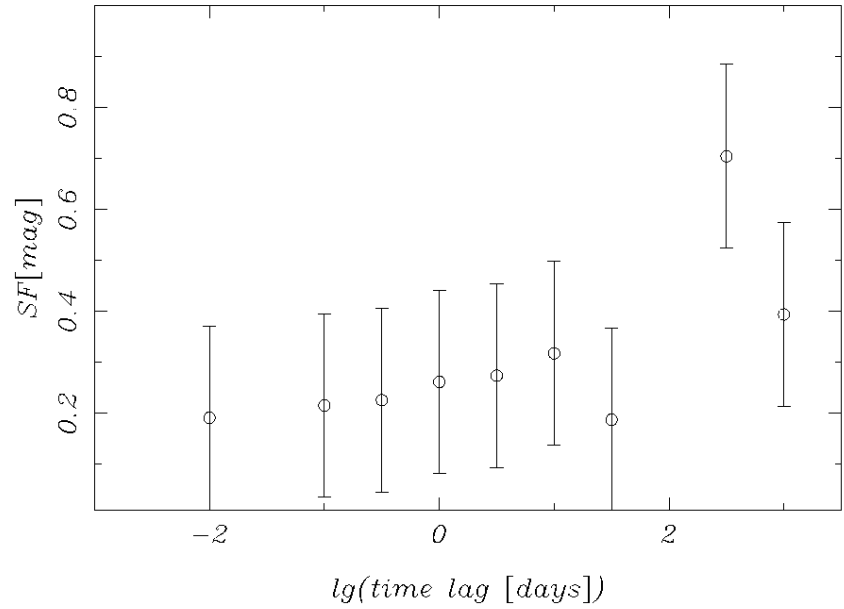,width=4.4in,angle=270}
}
\caption{
B-band structure function of $\sim500$ quasars for $1 < z < 4$
using PQ and SDSS data. We are extending this to tens of thousands of quasars.
}
\end{figure}

\begin{figure} 
\centerline{
\psfig{file=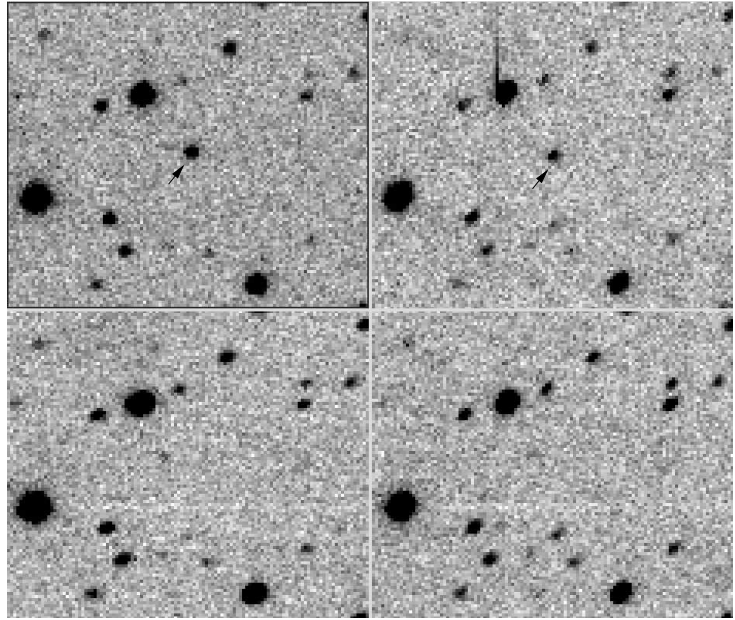,width=3.8in,angle=0}
}
\caption{
Example of an asteroid from $\it{R}$ and $\it{I}$ images from Aug and Jun 2004.
The object easily passed as a star. Only after 
using the known asteroid database could it be identified with one. Faster
asteroids are often easier to track as they form an elongated image.
}
\end{figure}

\section{Detecting Transients in Real Time}

The existing {\it PQ} pipeline is geared to complete processing of a
night's data by the next day.
In a matter of hours catalogs become available in the four filters used and can
be compared with other PQ epochs, or with other sky surveys and pointings for
the area covered that night.
This is sufficient for most {\it PQ} projects, including those involving
variable objects, e.g. SNe.  However, 
for the follow-up of rapidly fading sources and transients a faster
turn-around time is needed. Thus, we have
started work on a real-time processing pipeline which will enable detections
of such sources within minutes or less.
The following steps will be part of the process:
(1) Compare nightly catalogs with older catalogs from {\it PQ} itself,
(2) Compare the catalogs with other surveys and archives, using NVO
infrastructure and methodology,
(3) Compare positions of possible transients with those of known
variable sources, known asteroids, etc. (see Figure 2 for an example),
(4) A source classification engine will use the catalog data and the image
data to categorize the possible transients to determine the likely types,
(5) A decision engine will determine the significance of the object: should it be 
broadcast to the email subscribers list so that it can be followed up immediately,
or should it go to the website to be announced as a lower priority but 
interesting transient, or just flagged as an interesting object to be
looked at again in future epochs (in addition to other categories like
possible variable, orphan GRB etc.)

Given the large area coverage of 
{\it PQ}
and the results from our exploratory DPOSS project noted above, we estimate
that we will be detecting up to several tens of highly variable or transient 
sources per night.
Points (4) and (5) above summarize the key challenge of dealing with this abundance
of data in an effective
manner -- maintaining a high completeness in terms of the interesting variable
and transient sources discovered, while maintaining a low contamination rate
by spurious or uninteresting sources -- and doing it in real time.
We will utilize distributed Grid based services developed by the GRIST
project (Jacob et al. 2005) along with a variety of
advanced statistical (Mahabal et al. 2004, Graham et al. 2005) and Machine
Learning techniques towards this end.

\vskip0.2in
\noindent{\bf Acknowledgments:}  This work was supported in part by the
NSF grants AST-0326524, AST-0407448 and the NASA contract NAG5-9482.
PK and EK were supported in part by SURF Fellowships at Caltech.
SGD acknowledges a partial support from the Ajax Foundation.

\end{document}